\documentclass{article}
\usepackage{graphicx}

\input{tcilatex}
\begin{document}

\begin{center}
\underline{{\Large Regulaci\'{o}n de especies competitivas bajo modelos
impulsivos}}

\underline{{\Large \ de pesca-siembra regidos por operadores maximales}}

\bigskip 

Romina Cardo \& Alvaro Corval\'{a}n

{\small rcardo@ungs.edu.ar, acorvala@ungs.edu.ar}

\bigskip 

Instituto del Desarrollo Humano. Universidad Nacional de General Sarmiento
\end{center}

\bigskip

INTRODUCCI\'{O}N

Las ecuaciones diferenciales impulsivas se han comenzado a usar
recientemente con \'{e}xito para modelar el comportamiento de la biomasa de
poblaciones individuales ante esquemas de captura pulsante masiva y per\'{\i}%
odos de veda. (ver p. ej [2])

En ciertos \'{a}mbitos lacustres en que se permite de forma habitual la
pesca deportiva tiene lugar un esquema distinto ya que la pesca se realiza
de manera continua, y se incrementa en forma impulsiva la biomasa de una
especie mediante la siembra masiva de alevines de la misma. (Los pescadores
prefieren fuertemente a las truchas debido a la mayor resistencia que oponen)

En los casos en que interact\'{u}an dos o m\'{a}s especies competitivas bajo
captura la din\'{a}mica se complica considerablemente. En muchos lagos patag%
\'{o}nicos abiertos a la pesca deportiva se produce dicha competencia entre
especies aut\'{o}ctonas (existentes en la regi\'{o}n incluso en el registro f%
\'{o}sil), por un lado: las percas (Percichthys trucha --o trucha criolla, o
perca de boca chica-, y Percichthys colhuapiensis --perca de boca grande-),
y las especies ex\'{o}ticas: trucha arco\'{\i}ris (Oncorhynchus mykiss) y
trucha marr\'{o}n (Salmo trutta), introducidas recientemente por motivos
comerciales. La pr\'{a}ctica de la siembra intensiva ocasional de alevines
de truchas introduce variaciones pulsantes abruptas que f\'{a}cilmente
pueden inducir una din\'{a}mica ca\'{o}tica(ver [3]).

Como veremos, en determinadas condiciones las especies aut\'{o}ctonas llegan
a veces a puntos cercanos a la extinci\'{o}n en determinado habitat. Si bien
es posible reintroducirlas -aunque rara vez se hace-, no es lo ideal, ya que
va en sentido contrario a la preservaci\'{o}n de la diversidad gen\'{e}tica
de dichas epecies.

En otros casos el efecto de las siembras es menor al deseado, y la poblaci%
\'{o}n de truchas tiende a disminuir en promedio, a pesar de los impulsos.

En este trabajo consideramos una estrategia para la decisi\'{o}n de los
momentos de siembra teniendo en cuenta el valor del operador maximal de
Hardy-Littlewood (a izquierda) para la masa de la poblaci\'{o}n a
introducir; dicha estrategia puede presentarse como natural desde el punto
de vista de quienes propulsan la siembra, y por otra parte podemos obtener
en esta situaci\'{o}n resultados que permiten controlar las magnitudes
resultantes de las especies.

Aqu\'{\i} consideramos ecosistemas relativamente cerrados (estanques,
lagunas, lagos) donde puedan despreciarse las migraciones, en latitudes al
sur del paralelo 40 S -v.g., Lago Argentino, Laguna Verde, Lago Strobel,
Lago Largo, Lago Futalaufqu\'{e}n, etc. En otras provincias argentinas: C%
\'{o}rdoba, La Pampa, Buenos Aires, suele realizarse tambi\'{e}n siembra de
truchas pero all\'{\i} el problema no se presenta debido a que en ellas hay
per\'{\i}odos del a\~{n}o en que las temperaturas relativamente altas, lo
cual disminuye la cantidad de ox\'{\i}geno disuelto en el agua. A partir de
los 17/18%
${{}^o}$
C las truchas tienen dificultades para digerir el alimento por carencia de ox%
\'{\i}geno, y temperaturas superiores a 20%
${{}^o}$
C resultan letales para las truchas.[4]

Por simplicidad consideraremos el escenario donde interact\'{u}an 2 grupos
de peces, ya sea porque s\'{o}lo hay una especie aut\'{o}ctona de inter\'{e}%
s y s\'{o}lo 1 ex\'{o}tica, o bien reuniendo por un lado las poblaciones aut%
\'{o}ctonas y por otro las for\'{a}neas: $X\left( t\right) =\left( 
\begin{array}{c}
X_{1}\left( t\right) \\ 
X_{2}\left( t\right)%
\end{array}%
\right) $, donde $X_{1}$ es la biomasa del grupo originario (Percas), y $%
X_{2}$ corresponde a los peces ex\'{o}ticos (Truchas).

\bigskip

MODELOS\ IMPULSIVOS\ CON\ TIEMPOS\ DE\ IMPULSOS\ PREFIJADOS

En las ecuaciones diferenciales impulsivas se consideran usualmente dos
escalas de tiempo, una continua donde la evoluci\'{o}n es regida por una
ecuaci\'{o}n diferencial, y otra discreta donde para una secuencia de
instantes aislados la evoluci\'{o}n es impulsiva:

\[
\left\{ 
\begin{array}{c}
X^{\prime }(t)=f(t,X(t))\text{ si }t\neq \tau _{k} \\ 
X(t^{+})=g(k,X)\text{ si }t=\tau _{k}%
\end{array}%
\right. 
\]

Habitualmente los tiempos $\{\cdots ,\tau _{-2},\tau _{-1},\tau _{0},\tau
_{1},\tau _{2},\cdots \}$ de impulso son prefijados (equiespaciados o no) y
los impulsos $X\left( \tau _{k}^{+}\right) $ dependen de $k$ y s\'{o}lo del
valor instant\'{a}neo de $X$ (es decir $X\left( \tau _{k}^{+}\right) $).

\bigskip

MODELO\ IMPULSIVO\ CON\ TIEMPOS\ DE\ IMPULSOS

DEPENDIENTES\ DE\ LOS$\ $VALORES\ HIST\'{O}RICOS DE $X$

Consideramos en cambio un modelo impulsivo en que los tiempos de impulso
dependen de la imagen por $X$ de los tiempos previos -as\'{\i} como tambi%
\'{e}n las magnitudes de los impulsos.

\[
\left\{ 
\begin{array}{c}
X^{\prime }(t)=f(t,X(t))\text{ si }t\neq \tau _{k} \\ 
X(t^{+})=g(K,X\left( (-\infty ,\tau _{k}]\right) )\text{ si }t=\tau
_{k}=\inf \{t>\tau _{k-1}:Q(X\left( (-\infty ,t]\right) )>\lambda \}%
\end{array}%
\right. 
\]

MODELO\ CONCRETO\ CORRESPONDIENTE

AL\ PROBLEMA\ DE\ PESCA\_SIEMBRA

En nuestro problema modelamos la situaci\'{o}n en la escala continua por un
modelo de competencia log\'{\i}stico en la biomasa total de ambas especies
competidoras, donde $K$ indica el valor de saturaci\'{o}n, $q$ un factor que
indica la tasa de pesca, $P$ un vector que indica el porcentaje no devuelto
de las captura y $C$ la matriz que indica las influencias mutuas de las
especies (la dependencia de $X$ es solo a fin de la posible consideraciones
de masas cr\'{\i}ticas para cada especie, debajo de las cuales cabe esperar
que se extingan aunque sean no nulas -por ejemplo por escasas probabilidades
de cruzamiento-). El segundo t\'{e}rmino podr\'{\i}a incorporarse al
primero, modificando $C$, pero es preferible considerarlo aparte para
facilitar la interpretaci\'{o}n. Los valores de $C,$ $q$ y $P$ fueron
tomados a partir de datos del comportamiento conocido de las especies, datos
informados de pesca y el relevamiento encuestado del aprovechamiento de los
agentes respecto de las presas capturadas.

\[
X^{\prime }(t)=C\left( X(t)\right) \ast X(t)\ast \left( 1-\dfrac{\left\Vert
X\left( t\right) \right\Vert _{1}}{K}\right) -q\ast \left\langle P,X\left(
t\right) \right\rangle 
\]

Como las masas de las especies, y la contribuci\'{o}n de las mismas a la
derivada son estimativas, y adem\'{a}s pueden sufrir peque\~{n}as
perturbaciones de diverso origen en realidad consideramos:

\[
X^{\prime }(t)=C\left( \widetilde{X}(t)\right) \ast \widetilde{X}(t)\ast
\left( 1-\dfrac{\left\Vert \widetilde{X}\left( t\right) \right\Vert _{1}}{K}%
\right) -q\ast \left\langle P,\widetilde{X}\left( t\right) \right\rangle 
\]

donde $\widetilde{X}=X+\eta $, siendo $\eta $ un vector aleatorio
distribuido normalmente -cuya norma es relativamente peque\~{n}a respecto de 
$X$.

Luego, el modelo impulsivo cl\'{a}sico con impulsos equiespaciados ser\'{a}:

\[
\left\{ 
\begin{array}{c}
X^{\prime }(t)=C\left( \widetilde{X}(t)\right) \ast \widetilde{X}(t)\ast
\left( 1-\dfrac{\left\Vert \widetilde{X}\left( t\right) \right\Vert _{1}}{K}%
\right) -q\ast \left\langle P,\widetilde{X}\left( t\right) \right\rangle 
\text{ si }t\neq \tau _{k} \\ 
X(t^{+})=g(k,X)\text{ si }t=\tau _{k}%
\end{array}%
\right. 
\]

Como mencionamos, el \'{e}xito de las pol\'{\i}ticas de siembra pulsante
suele ser relativamente contingente, y frecuentemente ocurren situaciones de
inestabilidad no deseadas, donde tiene lugar, a la larga, la prevalencia de
una u otra especie, y esto para distintas tasas de siembra, y distintas
frecuencias para la escala de tiempo impulsiva.

Esto se ha observado emp\'{\i}ricamente en casos reales, y tambi\'{e}n puede
notarse en modelos simulados, por ejemplo:

\begin{center}
\FRAME{fhF}{2.6965in}{2.0323in}{0pt}{}{}{old25dp.eps}{\special{language
"Scientific Word";type "GRAPHIC";maintain-aspect-ratio TRUE;display
"PICT";valid_file "F";width 2.6965in;height 2.0323in;depth
0pt;original-width 5.7519in;original-height 4.3284in;cropleft "0";croptop
"1";cropright "1";cropbottom "0";filename '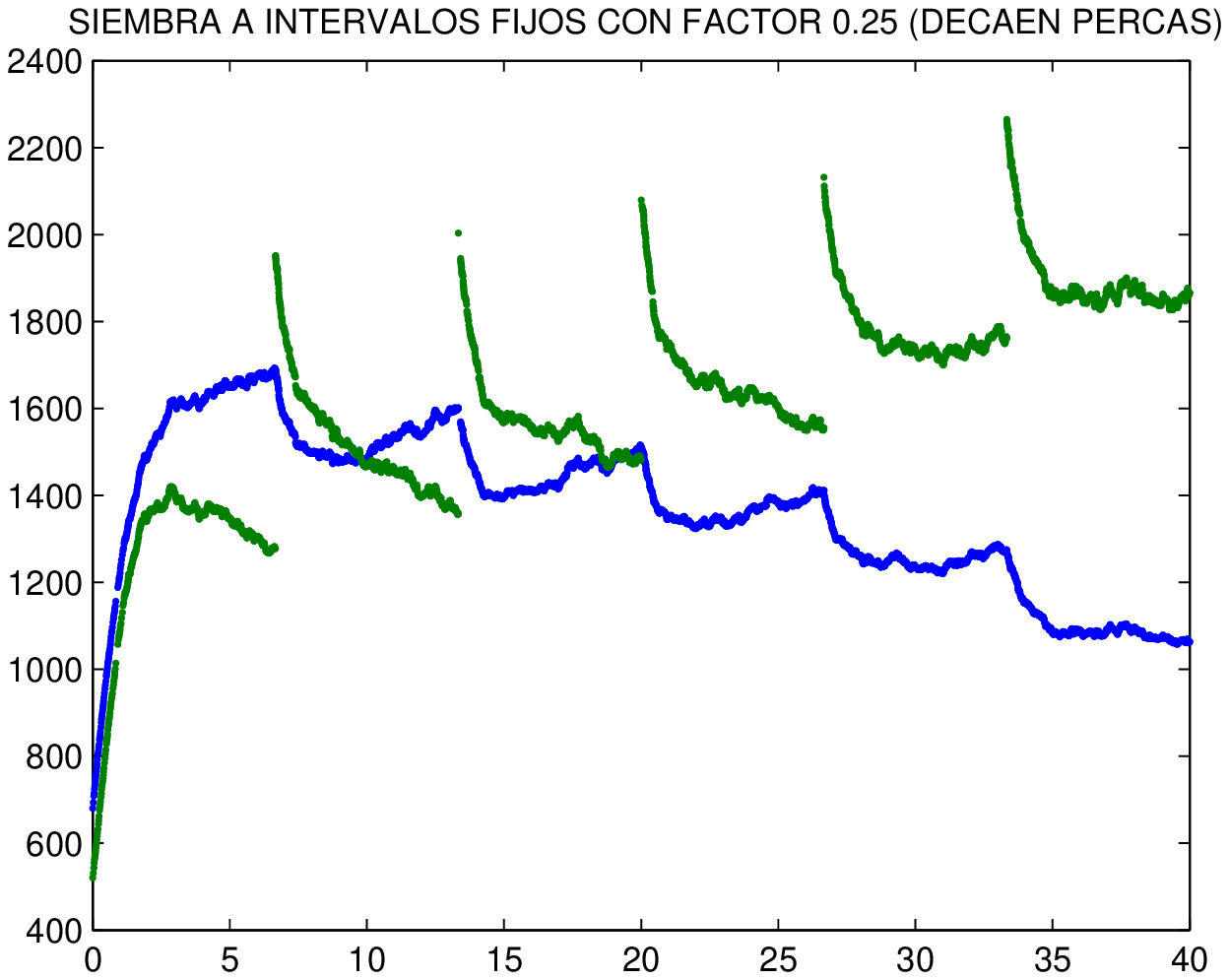';file-properties
"XNPEU";}}\FRAME{fhF}{2.6775in}{2.0176in}{0pt}{}{}{old25dt.eps}{\special%
{language "Scientific Word";type "GRAPHIC";maintain-aspect-ratio
TRUE;display "PICT";valid_file "F";width 2.6775in;height 2.0176in;depth
0pt;original-width 5.8219in;original-height 4.3708in;cropleft "0";croptop
"0.9830";cropright "0.9811";cropbottom "0";filename
'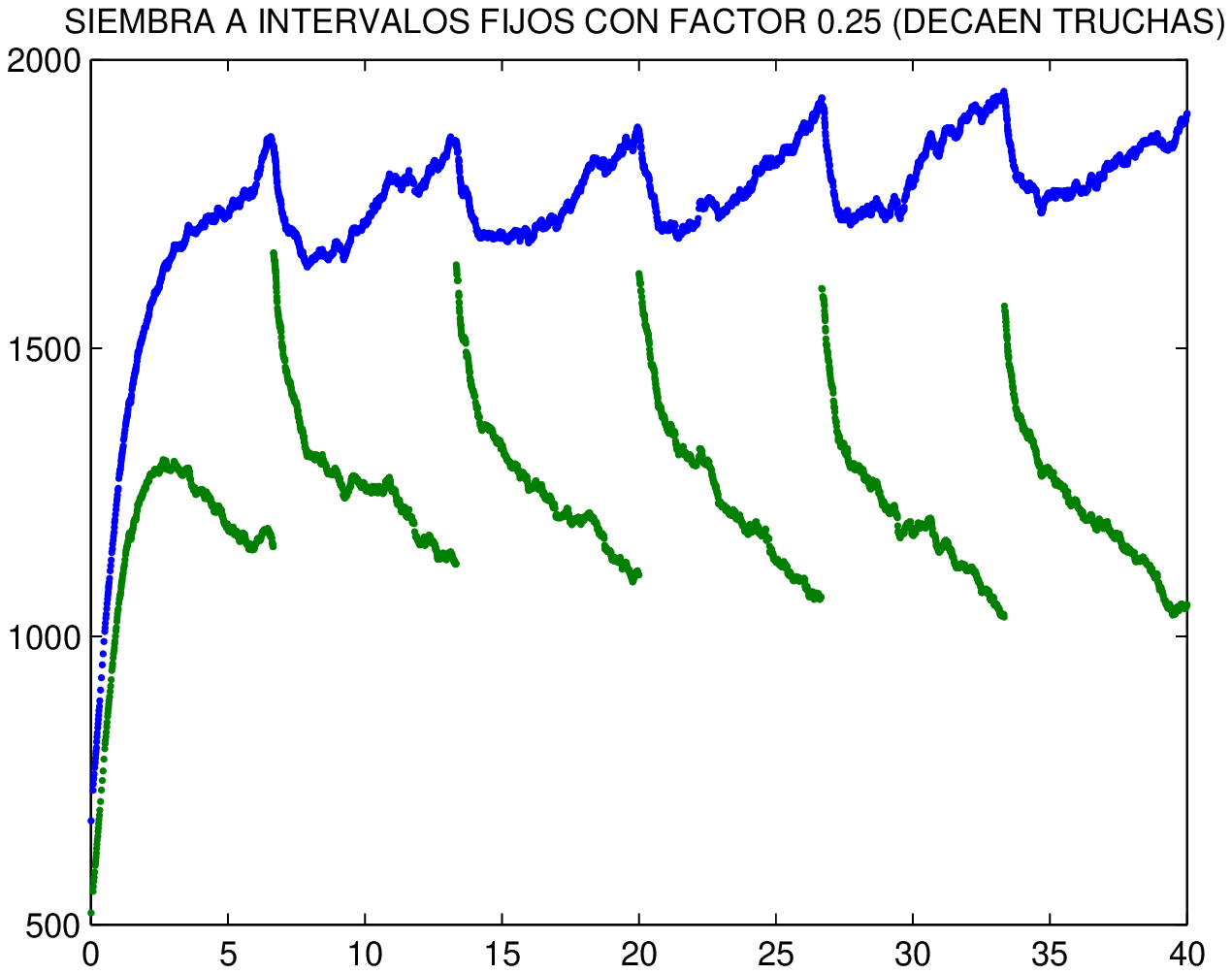';file-properties "XNPEU";}}

\FRAME{fhF}{2.7709in}{2.0851in}{0pt}{}{}{old35dp.eps}{\special{language
"Scientific Word";type "GRAPHIC";maintain-aspect-ratio TRUE;display
"PICT";valid_file "F";width 2.7709in;height 2.0851in;depth
0pt;original-width 5.8366in;original-height 4.3708in;cropleft "0";croptop
"1";cropright "0.9968";cropbottom "0";filename '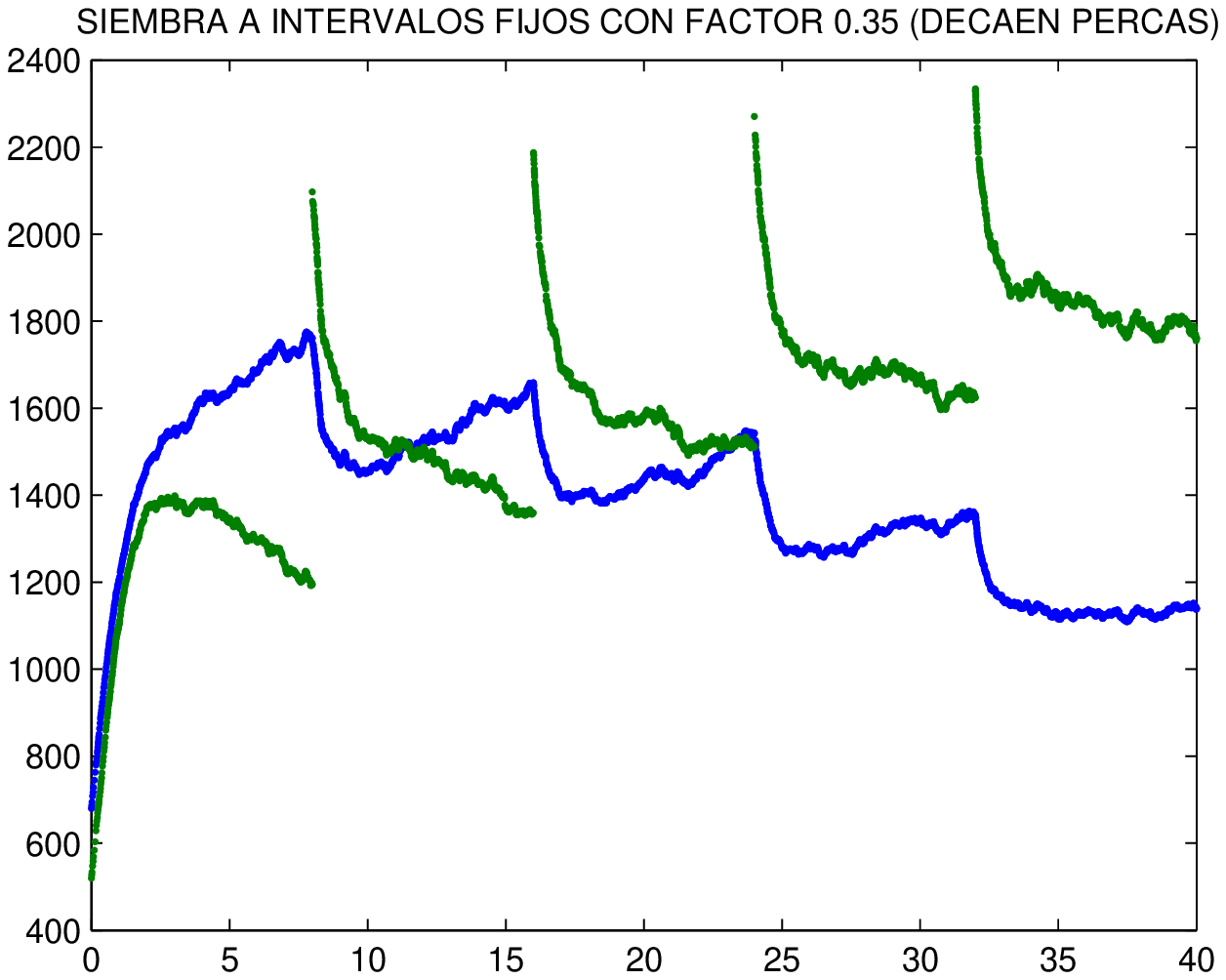';file-properties
"XNPEU";}}\FRAME{fhF}{2.7458in}{2.0721in}{0pt}{}{}{old35dt.eps}{\special%
{language "Scientific Word";type "GRAPHIC";maintain-aspect-ratio
TRUE;display "PICT";valid_file "F";width 2.7458in;height 2.0721in;depth
0pt;original-width 5.8219in;original-height 4.3708in;cropleft "0";croptop
"1.0034";cropright "1";cropbottom "0";filename 'OLD35DT.eps';file-properties
"XNPEU";}}

\FRAME{fhF}{2.725in}{2.0557in}{0pt}{}{}{old45dp.eps}{\special{language
"Scientific Word";type "GRAPHIC";maintain-aspect-ratio TRUE;display
"PICT";valid_file "F";width 2.725in;height 2.0557in;depth 0pt;original-width
5.8219in;original-height 4.3708in;cropleft "0";croptop "1.0031";cropright
"1";cropbottom "0";filename '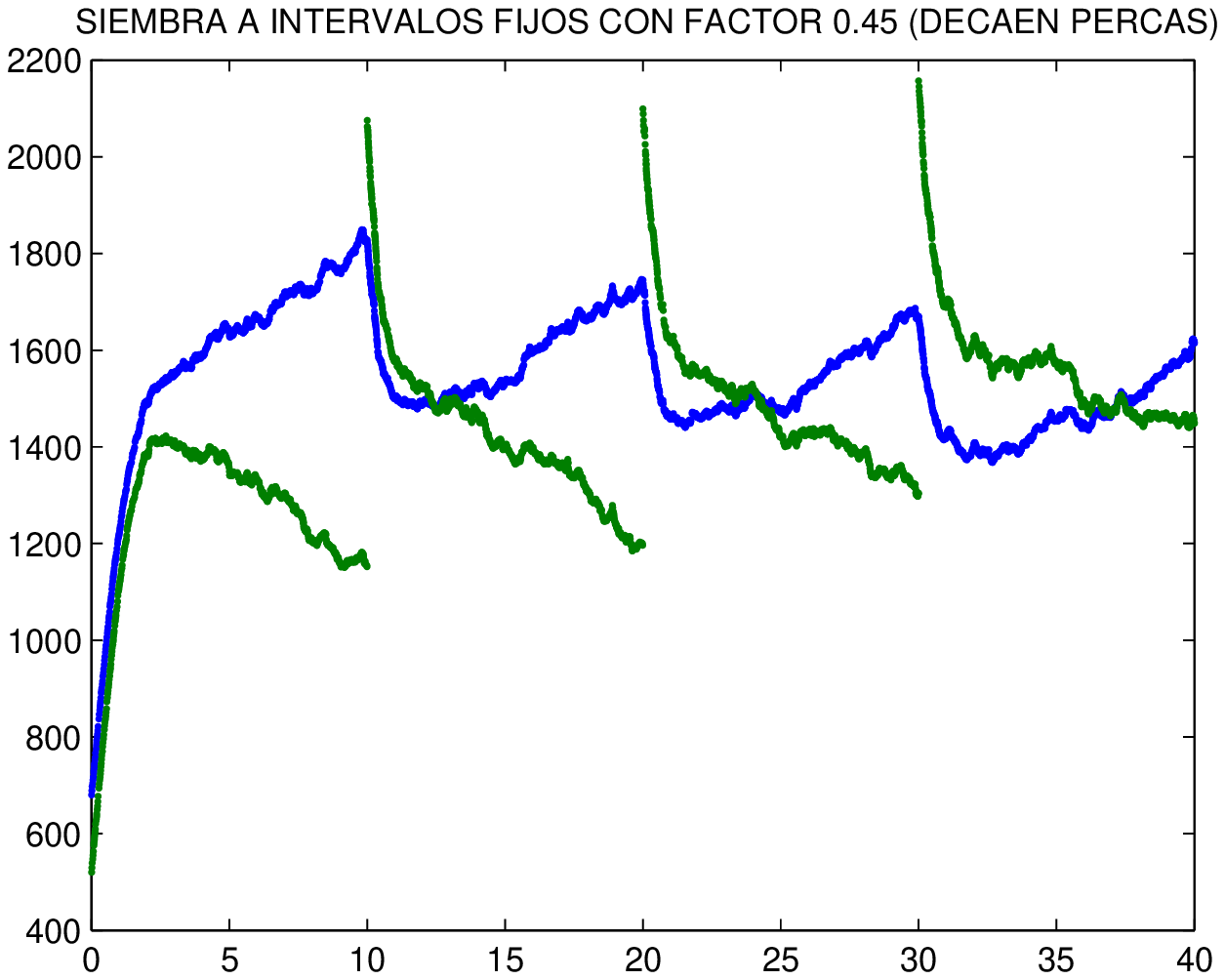';file-properties "XNPEU";}}\FRAME{%
fhF}{2.6982in}{2.0634in}{0pt}{}{}{old45dt.eps}{\special{language "Scientific
Word";type "GRAPHIC";display "PICT";valid_file "F";width 2.6982in;height
2.0634in;depth 0pt;original-width 5.8219in;original-height 4.3708in;cropleft
"0";croptop "1";cropright "0.9984";cropbottom "0";filename
'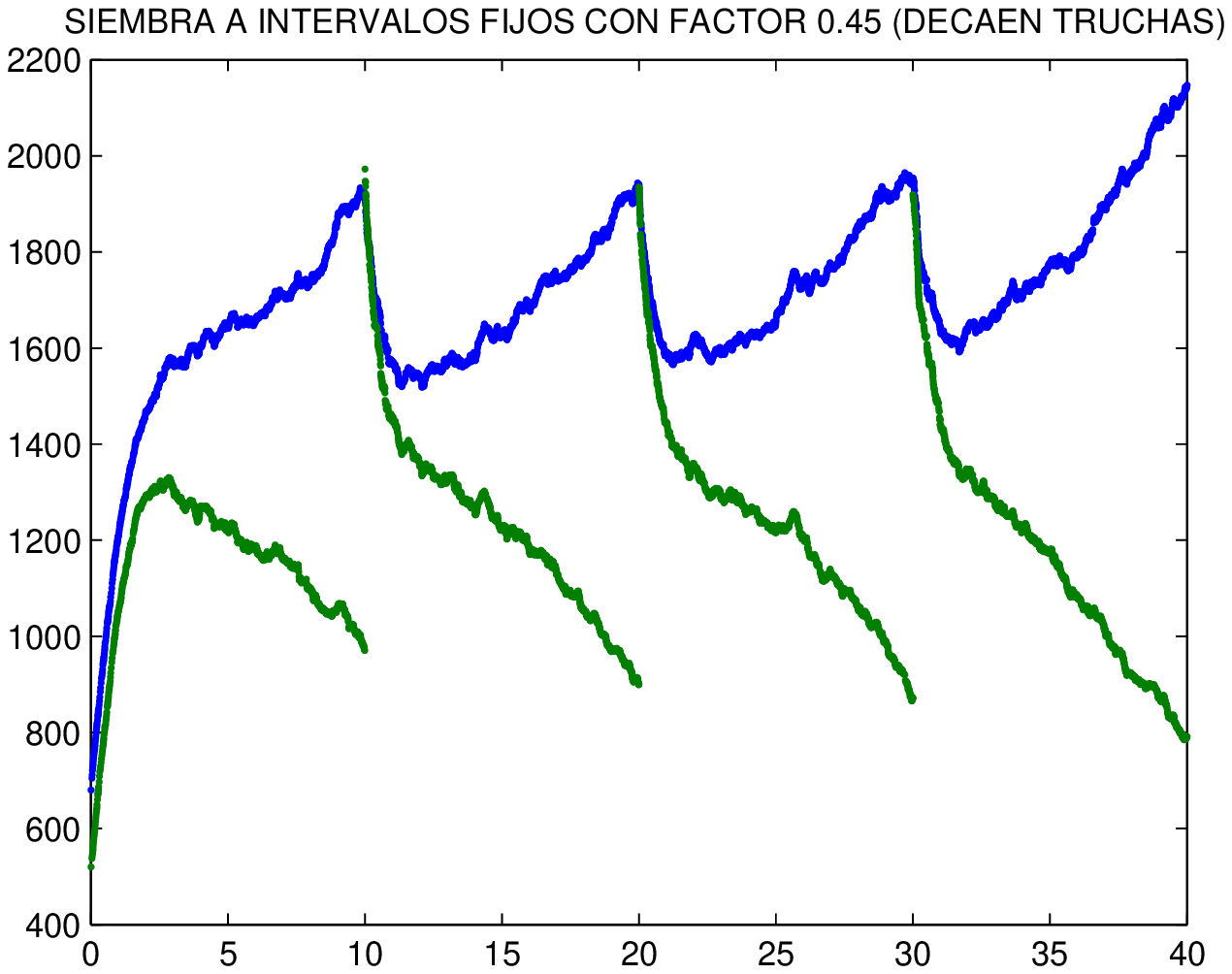';file-properties "XNPEU";}}
\end{center}

\bigskip

MODELO\ CON TIEMPOS\ DE\ IMPULSOS DEPENDIENDO

DE VALORES DE\ OPERADORES\ SOBRE $X\left( (-\infty ,t]\right) $

Teniendo en cuenta lo anterior hemos considerado una estrategia de siembra
determinada por los valores de las biomasas.

\[
\left\{ 
\begin{array}{c}
X^{\prime }(t)=C\left( \widetilde{X}(t)\right) \ast \widetilde{X}(t)\ast
\left( 1-\dfrac{\left\Vert \widetilde{X}\left( t\right) \right\Vert _{1}}{K}%
\right) -q\ast \left\langle P,\widetilde{X}\left( t\right) \right\rangle 
\text{ si }t\neq \tau _{k} \\ 
X(t^{+})=g(K,X\left( (-\infty ,\tau _{k}]\right) )\text{ si }t=\tau
_{k}=\inf \{t>\tau _{k-1}:Q(X\left( (-\infty ,t]\right) )>\lambda \}%
\end{array}%
\right. 
\]

Es decir donde los impulsos (es decir las siembras) se realizan cuando una
cantidad dependiendo de los valores previos de las masas de las especies
supera una determinada cota de tolerancia $\lambda $. Y el valor del impulso
depende tambi\'{e}n de valores actuales o previos, y posiblemente de la
constante de saturaci\'{o}n del h\'{a}bitat.

Es conveniente considerar valores hist\'{o}ricos, y no puntuales, ya que una
medici\'{o}n aislada no proporciona suficiente informaci\'{o}n para estimar
la evoluci\'{o}n del sistema. En la pr\'{a}ctica, dos sistemas similares que
en un determinado momento registren iguales valores de $X_{1}$ y $X_{2}$ en
general no evolucionar\'{a}n de la misma manera.

\bigskip

Yendo al extremo, si en un instante dado se introducen en un estanque virgen
ambas especies en cierta proporci\'{o}n: $X\left( t_{0}\right) =\left( 
\begin{array}{c}
X_{1}\left( t_{0}\right) \\ 
X_{2}\left( t_{0}\right)%
\end{array}%
\right) $, no se comportar\'{a} igual que uno donde ambas especies conviven
desde hace tiempo y en la que en $t_{0}$ la medici\'{o}n tambi\'{e}n sea: $%
X\left( t_{0}\right) =\left( 
\begin{array}{c}
X_{1}\left( t_{0}\right) \\ 
X_{2}\left( t_{0}\right)%
\end{array}%
\right) $. Entre otras causas esto se debe a que la medici\'{o}n releva
involucra ejemplares adultos y se subestima la contribuci\'{o}n de huevas,
alevines y ejemplares j\'{o}venes que m\'{a}s adelante influir\'{a}n en la
biomasa. La medici\'{o}n hist\'{o}rica permite tener en cuenta dicha
contribuci\'{o}n ya que las cantidades de ejemplares inmaduros depende de
ella.

Un modelo propuesto, en l\'{\i}nea con el inter\'{e}s de los lobbys de
pesca, es considerar las diferencias entre la masa total de las percas y la
de las truchas, y sembrar en el primer instante en que el mayor promedio de
una ponderaci\'{o}n de las diferencias, en per\'{\i}odos previos supera $%
\lambda $.

Esto equivale a considerar para $f\left( t\right)
=a_{1}X_{1}(t)-a_{2}X_{2}(t)$, el valor del operador lateral (a izquierda de
Hardy-Littlewood) $M^{-}$%
\[
M^{-}f\left( t\right) =\sup\limits_{h>0}\frac{1}{h}\dint\limits_{t-h}^{t}f%
\left( s\right) ds 
\]

y el criterio para el impulso ser\'{a}:%
\[
M^{-}f\left( t\right) =\sup\limits_{h>0}\frac{1}{h}\dint\limits_{t-h}^{t}f%
\left( s\right) ds>\lambda 
\]

Es interesante el criterio, y tiene ventajas para el an\'{a}lisis debido a
que es un operador de tipo d\'{e}bil $\left( 1,1\right) $. Por lo tanto si $%
\left\vert f\left( t\right) \right\vert $ tiende de manera razonable r\'{a}%
pida a $0$ (si las poblaciones tienden a equilibrar su diferencia en ciertos
valores de coexistencia) est\'{a} acotada la medida del conjunto de tiempos
en que $M^{-}f\left( t\right) $ supera $\lambda $, en funci\'{o}n de la
norma 1 de $f$, es decir .

En el caso del modelo de 2 especies competidoras -con siembra de alevines de
una especie preferida (entre las 2 existentes)- nuestro criterio para
decidir la siembra depende de si el valor de un operador maximal (a
izquierda, es decir en funci\'{o}n de valores del pasado) supera o no un
valor predeterminado $\lambda $.

Un criterio m\'{a}s conservador apuntar\'{\i}a a realizar los impulsos s\'{o}%
lo cuando los promedios hist\'{o}ricos desde el principio de las mediciones
(y no el promedio m\'{a}s favorable), a partir de un instante $0$
arbitrario. \ esto equivale a considerar el operador de Hardy $T$:

\[
Tf(t)=\frac{1}{t}\dint\limits_{0}^{t}f\left( s\right) ds 
\]

Tambi\'{e}n este es un operador acotado en $L^{1,\infty }$, y podr\'{\i}amos
considerar el criterio para el impulso:

\[
Tf\left( t\right) =\frac{1}{t}\dint\limits_{0}^{t}f\left( s\right)
ds>\lambda 
\]

Est\'{a} claro, en cualquier caso, que $M^{-}\geq T$, aunque, como dijimos,
ambos operadores son acotados en $L^{1,\infty }$, es decir que existe $%
C>0:\forall \lambda \geq 0$ 
\[
\left\vert \left\{ t:M^{-}\left\vert f\right\vert \left( t\right) >\lambda
\right\} \right\vert \leq C\left\Vert f\right\Vert _{1} 
\]%
C\'{o}mo $M^{-}\geq T$, la acotaci\'{o}n en $L^{1,\infty }$ de $M^{-}$
implica la de $T$. Aqu\'{\i} $\left\vert {}\right\vert $ representa la
medida de Lebesgue del conjunto. Pero tambi\'{e}n es posible probar la
acotaci\'{o}n en un contexto muy general, con pesos, y con otras medidas. (ve%
\'{a}se por ejemlo [5])

Considerando $f\left( t\right) =a_{1}X_{1}(t)-a_{2}X_{2}(t)$, tal que $%
f\left( t\right) \rightarrow 0$ cuando $t\rightarrow \infty $, es decir
siendo $\dfrac{a_{1}}{a_{2}}$ la relaci\'{o}n l\'{\i}mite entre las dos
especies competidoras si no hubiera perturbaciones, resulta que en este tipo
de modelo (de Levins, ver [1]), resulta $\left\Vert f\right\Vert _{1}<\infty 
$

\bigskip

En la pr\'{a}ctica se pueden encuentrar objeciones a ambos criterios:

Consideramos, en ambos casos los valores discretizados, con $a_{j}=f(j)$:

\begin{eqnarray*}
M^{-}f\left( n\right) &=&\sup\limits_{n\geq k\geq 1}\left( \frac{1}{n-k+1}%
\dsum\limits_{j=k}^{n}a_{j}\right) \\
Tf(t) &=&\frac{1}{n}\dsum\limits_{j=1}^{n}a_{j}
\end{eqnarray*}

Vayamos ahora a la cuesti\'{o}n de qu\'{e} es lo que tiene que suceder para
que el valor de uno de estos operadores supere un valor prefijado $\lambda $
en un instante $n+1$, si hasta el instante $n$ el valor del operador estaba
por debajo de $\lambda $:

En el caso del operador maximal la cuesti\'{o}n depende fuertemente del
nuevo valor $a_{n+1}$. En efecto, supongamos 
\[
\lambda \geq M^{-}f\left( n\right) =\frac{1}{n-k_{0}+1}\dsum%
\limits_{j=k_{0}}^{n}a_{j}\geq \frac{1}{n-k+1}\dsum\limits_{j=k}^{n}a_{j} 
\]%
(para cierto $k_{0}$ y para todo $k$). Ahora si $M^{-}f\left( n+1\right)
>\lambda $ tiene que ser 
\[
M^{-}f\left( n+1\right) =\frac{1}{\left( n+1\right) -k+1}\dsum%
\limits_{j=k}^{n+1}a_{j}>\lambda 
\]%
para cierto $k$ con $n+1\geq k\geq 1$, luego 
\[
\dsum\limits_{j=k}^{n+1}a_{j}>\left( \left( n+1\right) -k+1\right) \lambda
=\lambda +(n-k+1)\lambda 
\]%
, es decir 
\[
a_{n+1}+\dsum\limits_{j=k}^{n}a_{j}>\lambda +(n-k+1)\lambda 
\]%
, de donde: 
\[
a_{n+1}-\lambda >(n-k+1)\lambda -\dsum\limits_{j=k}^{n}a_{j} 
\]%
, por lo tanto: 
\[
\dfrac{a_{n+1}-\lambda }{(n-k+1)}>\lambda -\dfrac{1}{(n-k+1)}%
\dsum\limits_{j=k}^{n}a_{j}\geq \lambda -M^{-}f\left( n\right) \geq 0 
\]%
, y entonces $a_{n+1}-\lambda >0$, luego $a_{n+1}>\lambda $, que resulta una
condici\'{o}n necesaria para que $M^{-}f\left( n+1\right) >\lambda $ dado
que $\lambda \geq M^{-}f\left( n\right) $.

Pero adem\'{a}s, como $M^{-}f\left( n+1\right) =\sup\limits_{n+1\geq k\geq
1}\left( \frac{1}{\left( n+1\right) -k+1}\dsum\limits_{j=k}^{n+1}a_{j}%
\right) \ $resulta que el caso $k=n+1$ es menor que el caso general, es
decir: 
\[
M^{-}f\left( n+1\right) \geq \left( \frac{1}{\left( n+1\right) -(n+1)+1}%
\dsum\limits_{j=n+1}^{n+1}a_{j}\right) =\frac{1}{1}a_{n+1}=a_{n+1} 
\]%
, de modo que si $a_{n+1}>\lambda $ entonces $M^{-}f\left( n+1\right) \geq
a_{n+1}>\lambda $.

Luego $M^{-}f\left( n+1\right) >\lambda $. Es decir que si $\lambda \geq
M^{-}f\left( n\right) $, entonces $a_{n+1}>\lambda $ es condici\'{o}n
necesaria y suficiente para que $M^{-}f\left( n+1\right) >\lambda $.

Es decir, que en el caso discreto el criterio $M^{-}f\left( n\right)
>\lambda $, no difiere de observar la diferencia ponderada puntual $f\left(
t\right) =a_{1}X_{1}(t)-a_{2}X_{2}(t)$ e impulsar siempre que supere cierto
valor $\lambda $. La condici\'{o}n es entonces bastante laxa (no depende de $%
n$).

Sin embargo uno \underline{quisiera} tener en cuenta el comportamiento de la
funci\'{o}n en per\'{\i}odos relativamente extensos del pasado y no s\'{o}lo
en el \'{u}ltimo valor puntual.

Consideremos por otra parte la situaci\'{o}n para el operador de Hardy $T$.
Si $\lambda \geq Tf\left( n\right) =\frac{1}{n}\dsum\limits_{j=1}^{n}a_{j}$,
para que 
\[
\frac{1}{n+1}\dsum\limits_{j=1}^{n+1}a_{j}=Tf\left( n+1\right) >\lambda 
\]%
necesitamos que $\dsum\limits_{j=1}^{n+1}a_{j}>\left( n+1\right) \lambda $,
o sea 
\[
a_{n+1}-\dsum\limits_{j=1}^{n}a_{j}>\lambda +n\lambda 
\]%
luego $a_{n+1}-\lambda >n\lambda -\dsum\limits_{j=1}^{n}a_{j}$, entonces: 
\[
a_{n+1}-\lambda >n\cdot (\lambda -\frac{1}{n}\dsum\limits_{j=1}^{n}a_{j}) 
\]

De aqu\'{\i} resulta que $a_{n+1}-\lambda >0$ a\'{u}n es condici\'{o}n
necesaria, pero en este caso ya no es suficiente. En t\'{e}rminos de la
suficiencia debe ser 
\[
a_{n+1}>\lambda +n\cdot (\lambda -Tf\left( n\right) ) 
\]%
que depende de $n$ y de la diferencia entre $\lambda $ y $Tf\left( n\right) $%
. Es decir que el $a_{n+1}$ debe ser mayor que $\lambda $ en una cantidad
igual a $n$ veces la diferencia entre $\lambda $ y $Tf\left( n\right) $. Por
lo tanto es cada vez m\'{a}s dif\'{\i}cil al aumentar $n,$ lo que a la larga
representa una condici\'{o}n muy restrictiva.

\bigskip

Decidimos entoncer usar como criterio que un promedio entre $M^{-}$ y $T$
supere $\lambda $, es decir $\alpha \cdot M^{-}f\left( n\right) +\left(
1-\alpha \right) \cdot Tf(n)>\lambda $, para balancear la puntualidad de $%
M^{-}$ y la globalidad de $T$.

\bigskip

Adem\'{a}s, como 
\[
\left\{ t:\alpha \cdot M^{-}f\left( t\right) +\left( 1-\alpha \right) \cdot
Tf\left( t\right) >\lambda \right\} \subset \left\{ t:\alpha \cdot
M^{-}\left\vert f\right\vert \left( t\right) >\lambda \right\} \cup \left\{
t:\left( 1-\alpha \right) \cdot Tf\left( t\right) >\lambda \right\} 
\]

se puede obtener la acotaci\'{o}n de $\alpha \cdot M^{-}+\left( 1-\alpha
\right) \cdot T$ a partir de las de $M^{-}$ y $T$.

\bigskip

Los valores de $a_{n+1}$ y de $\frac{1}{n+1}\dsum\limits_{j=1}^{n+1}a_{j}$
claramente no son independientes, pero la covarianza tiende a $0$ junto con $%
n$, de modo que si $Tf\left( n-1\right) $ es una fracci\'{o}n de $\lambda ,$
es decir $Tf\left( n-1\right) =\left( 1-\alpha \right) c\lambda $ con $0<c<%
\frac{1}{\left( 1-\alpha \right) }$, para que el criterio se dispare es
necesario, esencialemente, que $a_{n+1}>\alpha c\lambda $.

En las ensayos simulados, de hecho, con $\alpha =\frac{1}{2}$, es decir con
el criterio $\dfrac{M^{-}f\left( n\right) +Tf\left( n\right) }{2}>\lambda $,
o sea, promediando sencillamente las condiciones anteriormente consideradas
obtuvimos un desempe\~{n}o estable y bastante adecuado para distintos
valores tasas de siembra.

Cabe observar que, como dijimos, cada uno de los operadores est\'{a} acotado
en $L^{1,\infty }$ tambi\'{e}n el promedio lo est\'{a}, es decir existe $%
C>0:\forall \lambda \geq 0$ 
\[
\left\vert \left\{ t:\left( M^{-}+T\right) \left\vert f\right\vert \left(
t\right) >2\lambda \right\} \right\vert \leq C\left\Vert f\right\Vert _{1} 
\]

A continuaci\'{o}n mostramos algunos ejemplos de evoluciones, cuya
estabilidad era similar en varios centenares de ensayos (que inclu\'{\i}an
perturbaciones aleatorias normalmente distribuidas):

\begin{center}
\FRAME{fhF}{2.7458in}{2.0167in}{0pt}{}{}{new15.eps}{\special{language
"Scientific Word";type "GRAPHIC";display "PICT";valid_file "F";width
2.7458in;height 2.0167in;depth 0pt;original-width 5.8219in;original-height
4.3708in;cropleft "0";croptop "1";cropright "1";cropbottom "0";filename
'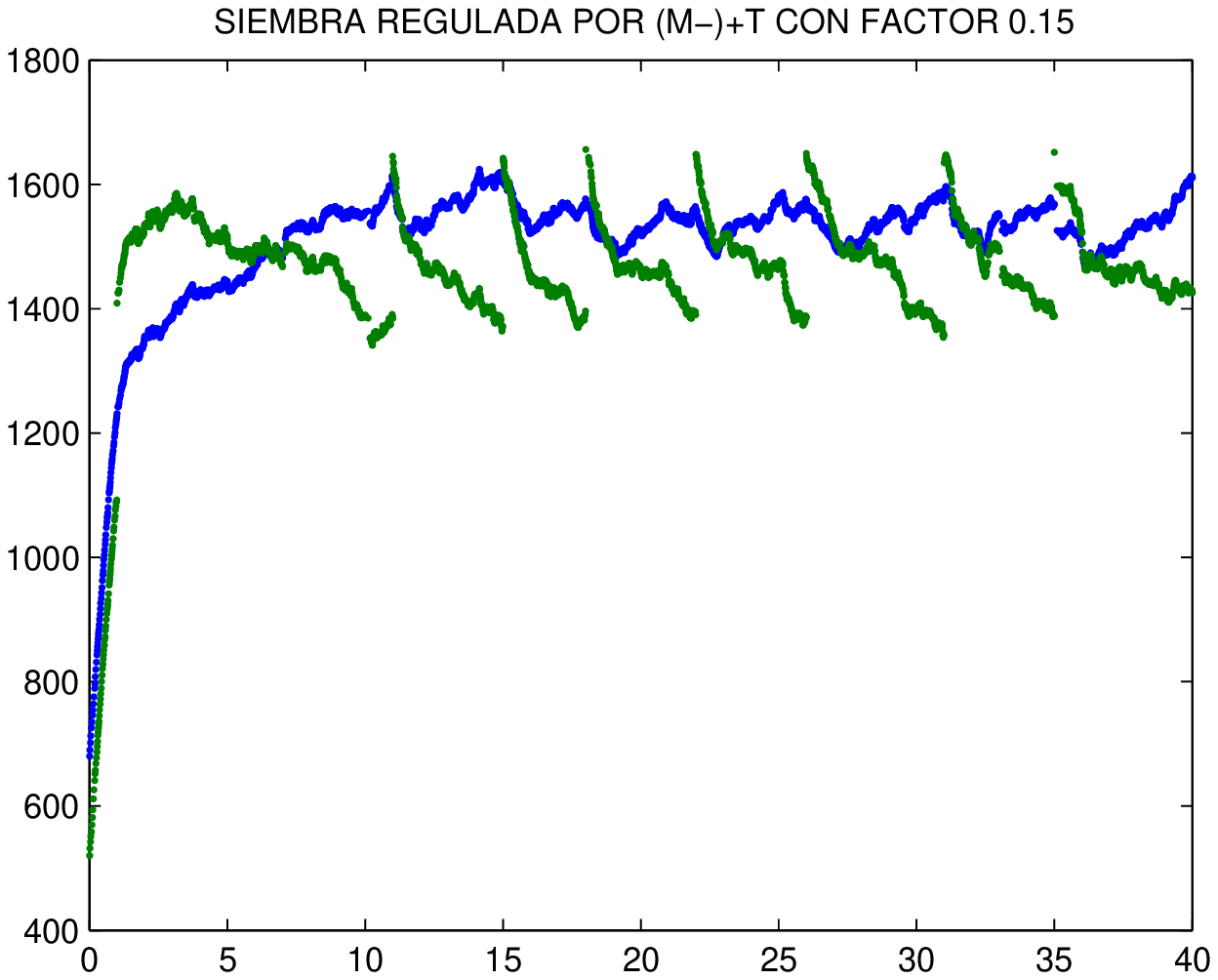';file-properties "XNPEU";}}\FRAME{fhFX}{2.7103in}{2.0211in}{0pt}{%
}{}{new25estable.eps}{\special{language "Scientific Word";type
"GRAPHIC";maintain-aspect-ratio TRUE;display "PICT";valid_file "F";width
2.7103in;height 2.0211in;depth 0pt;original-width 5.7519in;original-height
4.3284in;cropleft "0";croptop "0.9933";cropright "1.0039";cropbottom
"0";filename '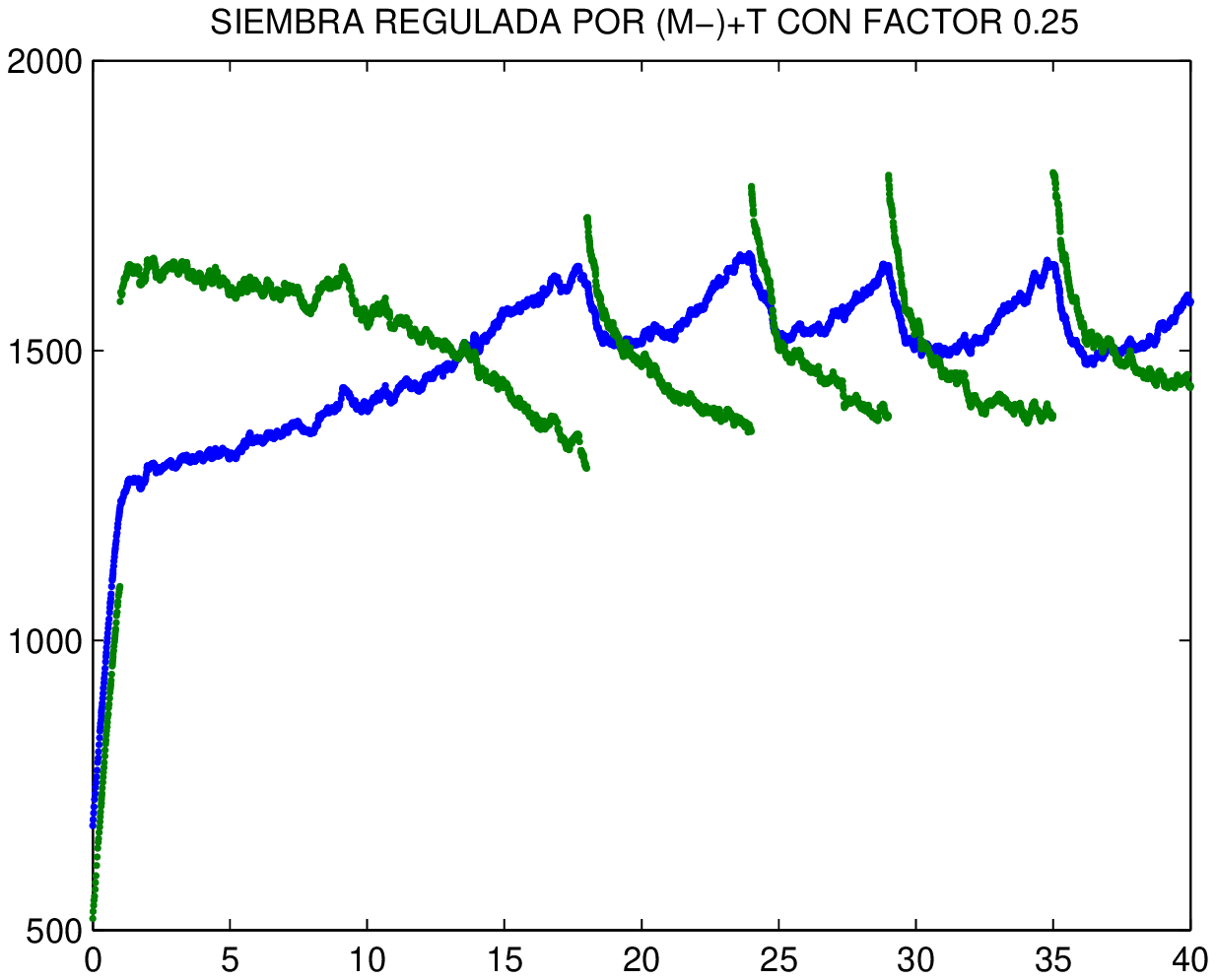';file-properties "XNPEU";}}

\FRAME{fhF}{2.7077in}{2.0176in}{0pt}{}{}{new35estable.eps}{\special{language
"Scientific Word";type "GRAPHIC";maintain-aspect-ratio TRUE;display
"PICT";valid_file "F";width 2.7077in;height 2.0176in;depth
0pt;original-width 5.8773in;original-height 4.3708in;cropleft "0";croptop
"1";cropright "1";cropbottom "0";filename '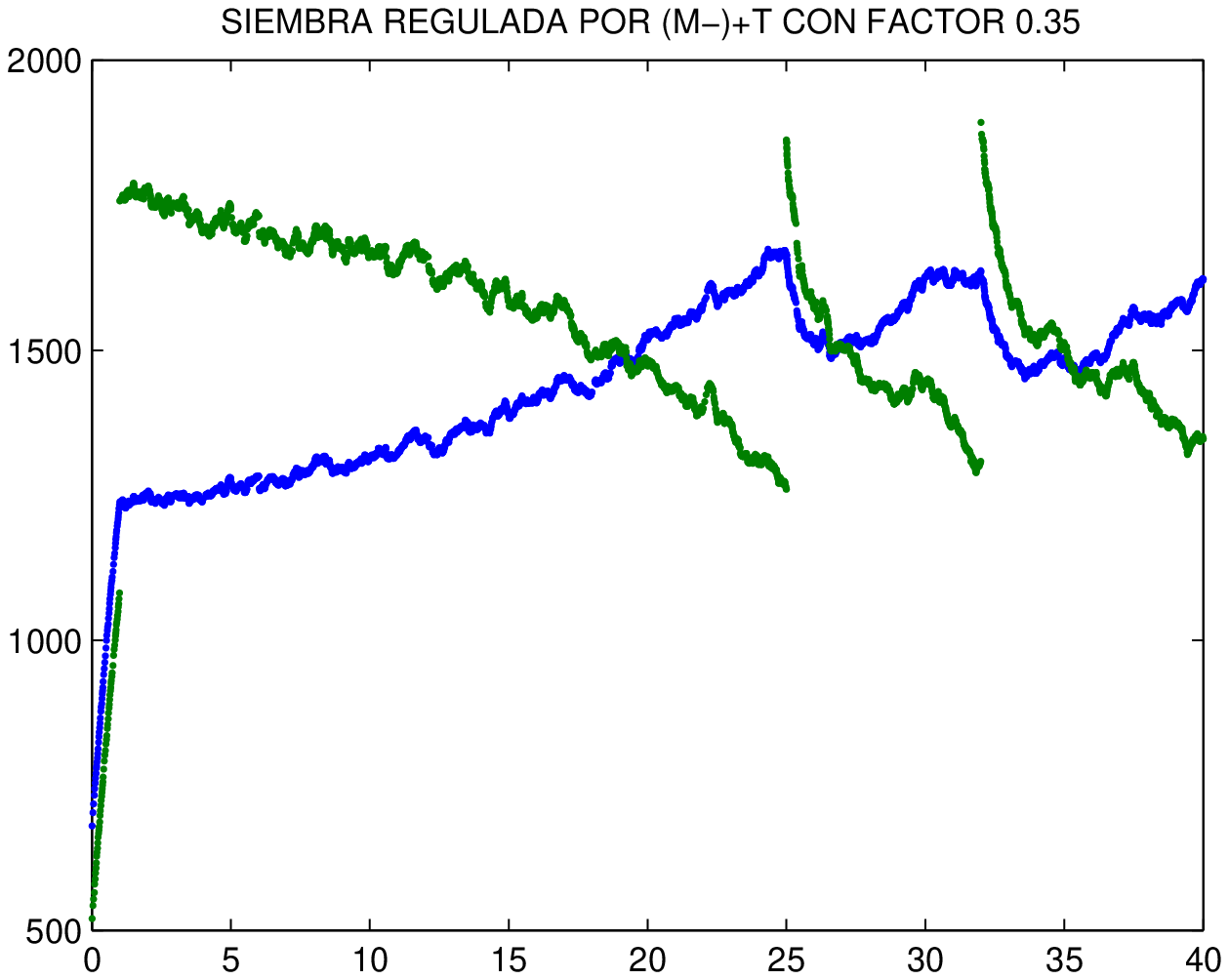';file-properties
"XNPEU";}}\FRAME{fbhF}{2.7562in}{2.0245in}{0pt}{}{}{new45.eps}{\special%
{language "Scientific Word";type "GRAPHIC";display "PICT";valid_file
"F";width 2.7562in;height 2.0245in;depth 0pt;original-width
5.8219in;original-height 4.3708in;cropleft "0";croptop "1";cropright
"1";cropbottom "0";filename '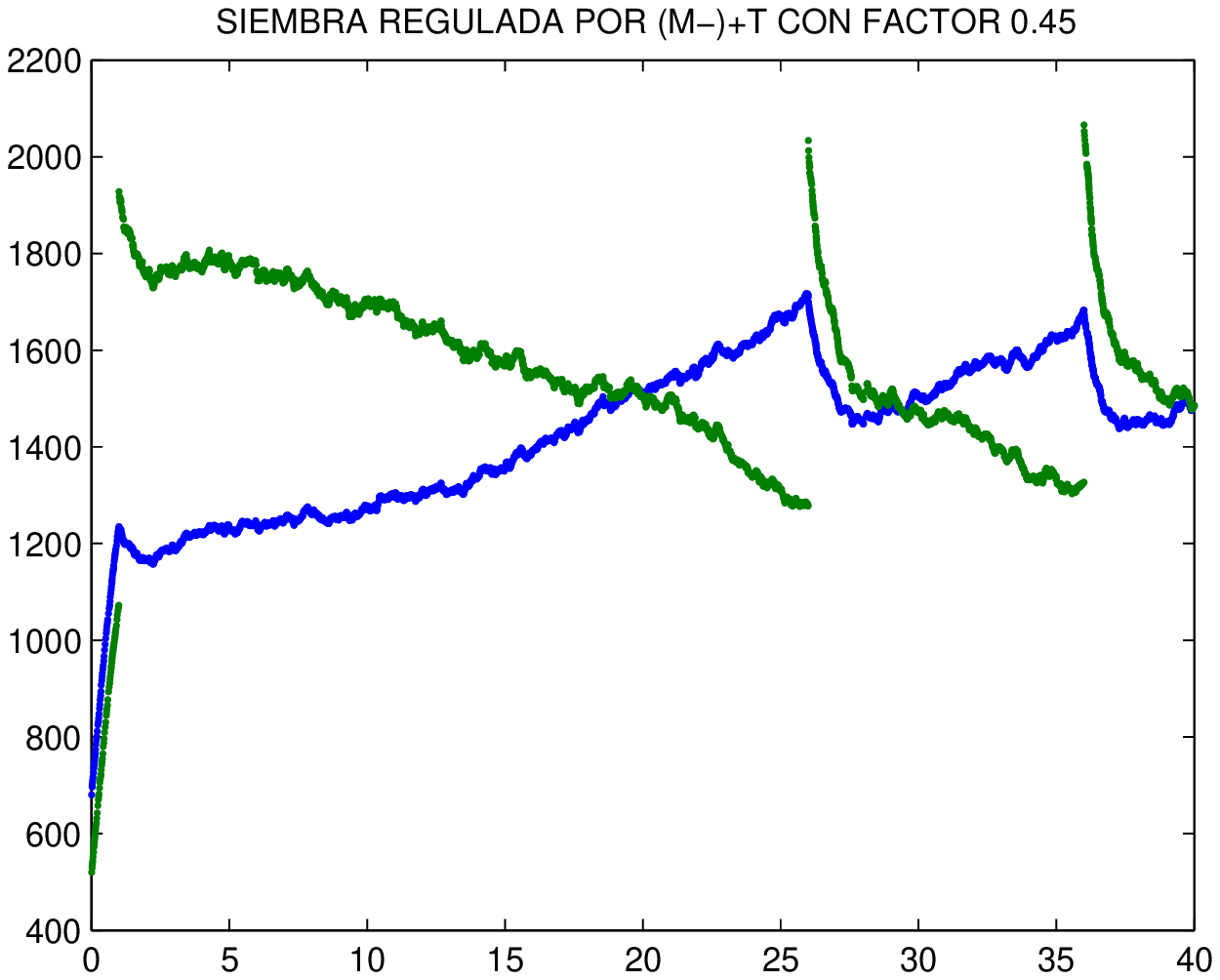';file-properties "XNPEU";}}
\end{center}

\bigskip

CONCLUSI\'{O}N PROVISIONAL

A modo de conclusi\'{o}n, los resultados de los ensayos sugieren que la
estrategia propuesta para la regulaci\'{o}n de los poblaciones siguiendo el
criterio de siembra determinado por los valores de $M^{-}+T$ mantendr\'{\i}a
una evoluci\'{o}n tendencialmente estable con distintas combinaciones de
criterio de impulso y factor de siembra, y manteniendo acotada la magnitud
total de las cantidades a introducir, permitiendo la pesca en los vol\'{u}%
menes y frecuencias habituales y evitando las propensiones a la extinci\'{o}%
n.

En principio, sobre los modelos num\'{e}ricos, los impulsos regidos por los
operadores propuestos tender\'{\i}an en general a la estabilidad, ya que los
impulsos llevar\'{\i}an los estados de un ciclo en el espacio de fases a
otro ciclo contenido en la c\'{a}psula convexa del anterior. En la pr\'{a}%
ctica, hay que tener en cuenta que las especies pueden extinguirse en un
determinado nicho ecol\'{o}gico en caso de que la cantidad de ejemplares
caiga debajo de cierto umbral cr\'{\i}tico y ser\'{a} conveniente obtener
datos de los resultados que se produzcan en el desempe\~{n}o de estas
estrategias en casos reales para sacar conclusiones definitivas.

\bigskip

\bigskip

BIBLIOGRAFIA

1. Abramson, Guillermo. "La matem\'{a}tica de los sistemas biol\'{o}gicos",
UNC (CNEA). 2013.

2. C\'{o}rdova Lepe, F., Del Valle, R. Robledo, G. "A pulse fishery model
with cIosures as function of the catch: Conditions for Sustainability.",
Mathematical Biosciences. Vol. 239, Issue 1, 2012

3. Bayo, Rodrigo E. "Comienzos de la truchicultura en la Provincia de Tierra
del Fuego", Asociaci\'{o}n Argentina de Acuicultura, 2013.

4. Municipalidad distrital de Ragash, Publicaciones del Centro de Estudios
para el desarrollo y la participaci\'{o}n. "Manual de Crianza de Truchas"
2009

5. Duoandikoetxea Zuazo, J. "Fourier Analysis", Graduate Studies in
Mathematics, American Mathematical Society, 2001.

\end{document}